\documentclass{aa}  

\usepackage{graphicx}
\usepackage{txfonts}
\begin{document}

%\title{Accurate frequency scales for customized wavelength calibrators}

\title{Accurate calibration spectra for precision radial velocities}
\subtitle{Iodine absorption referenced by a laser frequency comb}

%\titlerunning{Broadband iodine absorption spectrometry}

   \author{A. Reiners\inst{1}
     \and
     M. Debus\inst{1}
     \and
     S. Sch\"afer\inst{1}
     \and
     E. Tiemann\inst{2}
     \and
     M. Zechmeister\inst{1}
   }
   
   \institute{
     Institut f\"ur Astrophysik und Geophysik, Georg-August-Universit\"at, Friedrich Hund Platz 1, 37077 G\"ottingen, Germany
     \and
     Institut f\"ur Quantenoptik, Leibniz Universit\"at Hannover, Welfengarten 1, 30167 Hannover, Germany
   }
   
   \date{\today}

% \abstract{}{}{}{}{}
% 5 {} token are mandatory
 
  \abstract
  % context heading (optional)
  {Astronomical spectrographs require calibration of their dispersion
    relation, for which external sources like hollow-cathode lamps or
    absorption-gas cells are useful. Laser frequency combs (LFCs) are often
    regarded as ideal calibrators because they provide the highest accuracy and
    dense sampling, but LFCs are facing operational challenges such as
    generating blue visual light or tunable offset
    frequencies. %leave it empty if necessary
  % aims heading (mandatory)
    As an example of an external source, we aim to provide a precise and accurate frequency solution for the
    spectrum of molecular iodine absorption by referencing to an LFC that
    does not cover the same frequency range.
  % methods heading (mandatory)
    We used a Fourier Transform Spectrometer (FTS) to produce a consistent
    frequency scale for the combined spectrum from an iodine absorption cell at 5200--6200\AA\ and an LFC at 8200\,\AA. We used 17,807 comb lines to
    determine the FTS frequency offset and compared the calibrated iodine
    spectrum to a synthetic spectrum computed from a molecular potential
    model.
  % results heading (mandatory)
    In a single scan, the frequency offset was determined from the comb
    spectrum with an uncertainty of $\sim$1\,cm\,s$^{-1}$. The distribution of
    comb line frequencies is consistent with no deviation from linearity. The
    iodine observation matches the model with an offset of smaller than the model
    uncertainties of $\sim$1\,m\,s$^{-1}$, which confirms that the FTS zero
    point is valid outside the range covered by the LFC, and that the frequencies of  the iodine
    absorption model are accurate. We also report small
    systematic effects regarding the iodine model's energy scale.
    % conclusions heading (optional), leave it empty if necessary
    We conclude that Fourier Transform Spectrometry can transfer LFC accuracy
    into frequency ranges not originally covered by the comb. This allows us
    to assign accurate frequency scales to the spectra of customized
    wavelength calibrators. The calibrators can be optimized for individual
    spectrograph designs regarding resolution and spectral bandwidth, and
    requirements on their long-term stability are relaxed because FTS
    monitoring can be performed during operation.  This provides flexibility
    for the design and operation of calibration sources for high-precision
    Doppler experiments.  }

   \keywords{Molecular data -- Methods: laboratory: molecular --
     Instrumentation: spectrographs -- Techniques:
     radial velocities -- Reference systems}

   \maketitle
%
%________________________________________________________________

\section{Introduction}

Precise and accurate measurements of frequencies (or wavelengths) in
astronomical spectra enable a range of fundamental physical
experiments. Frequency shifts occur through Doppler shifts allowing the
measurement of velocities, which can be used, for example, to measure the mass
of unseen extrasolar planets \citep{2003A&A...401.1185L}, stellar pulsations
\citep{2004SoPh..220..137C}, and velocity fields in atmospheres of stars and
planets \citep{2000A&A...359..729A}. Transition frequencies of spectral lines
are determined and affected by fundamental constants like the fine structure
constant and the proton/electron mass ratio \citep{Dirac1937, Webb1999,
  2024ExA....57....5M}, by gravitational redshift \citep{Einstein1911,
  Einstein1916, 2003A&A...401.1185L}, and by the accelerated expansion of the
Universe, which affects galaxy motion on very large scales
\citep{2008MNRAS.386.1192L}. These and other experiments can be carried out if
the frequency scale in spectra from astronomical objects can be accurately
determined.

Astronomical spectroscopy is photon starved. Large telescopes are required to
collect enough light from distant objects for astrophysical analysis, which is
different from typical laboratory setups where the intensities of the investigated
light can often be controlled. For precision Doppler measurements, high spectral resolution is favorable \citep{2001A&A...374..733B}. \'Echelle
spectrographs reach resolutions of $R = \lambda / \Delta \lambda = 10^5$ at
efficiencies on the order of 10\,\%. For comparison, a Fourier Transform
Spectrometer (FTS) can operate at higher resolution and provides a number of
advantages regarding frequency calibration, but delivers an efficiency that is
several orders of magnitude lower than in astronomical spectrographs
\citep[see, e.g.,][]{2007ApJ...661..616H}.

In grating (\'echelle) spectrographs, light is collected in individual
detector pixels that are at minimum several hundred \,m\,s$^{-1}$ wide and are not
strictly evenly spaced \citep{2010MNRAS.405L..16W}. This poses a fundamental
problem to frequency calibration because, in principle, each individual pixel
requires calibration through external information. Furthermore, astronomical
spectroscopy often requires a relatively large bandwidth ---for example one
octave--- because information is collected simultaneously from many individual
spectral features. Calibration sources must provide dense and accurate
spectral information across the full frequency range of a spectrograph. Useful
calibration light sources are, for example, hollow cathode lamps
\citep{2008ApJS..178..374K}, absorption gas cells \citep{2010ApJ...713..410B},
Fabry-P\'erot etalons \citep[FPs;][]{2010SPIE.7735E..4XW, 2012SPIE.8446E..94S,
  2021AJ....161..252T, 2022arXiv221010988K, 10.1117/12.2629428,
  2022A&A...664A.191S, 2024MNRAS.530.1252S}, and laser frequency combs
\citep[LFC;][]{2008Sci...321.1335S, 2012Msngr.149....2L, 2019Optic...6..233M,
  10.1117/12.2624078, 10.1117/12.3019883}. This combination of requirements,
and especially the large wavelength range, is a challenge for the calibration sources
---including LFCs---  used for calibration in many high-precision
spectrographs \citep{2021A&A...646A.144S}. One of the advantages of an LFC is
that the frequency scale of its spectrum is accurately known from fundamental
principles, and that it provides a dense population of narrow lines, which
renders it a conceptually ideal reference. This is in contrast to hollow
cathode lamps where the distribution of lines is uneven, leaving large areas of
the spectral domain uncovered, and where individual lines are typically not
known to better than about 10\,m\,s$^{-1}$ \citep{Learner1988,
  2008ApJS..178..374K}. FPs can alleviate part of this problem by delivering a
tailored comb of peaks over a large wavelength range, but our knowledge of peak
frequencies and their stability is limited, necessitating external calibration
\citep{2015A&A...581A.117B}.

Gas-absorption cells are a spectroscopic standard for high-accuracy frequency
calibration, and are used, for example, in tunable laser applications
\citep[see][]{1978adsa.book.....G,
  GERSTENKORN1981322}. \citet{1979PASP...91..540C} introduced the use of
gas-absorption cells in astronomical observations, using hydrogen fluoride,
which was considered to be the most suitable available gas at the time. The
use of molecular iodine in astronomy dates back to observations of solar
Doppler shift measurements \citep{1983PhDT.........2K,
  1984A&A...134..134K}. Early applications used molecular absorption lines as
a reference for differential line shifts between the stellar (solar) and gas
absorption spectrum to track drifts in the spectral format. This is similar to
the use of telluric absorption lines as standard, which was introduced by
\citet{1973MNRAS.162..243G}, where the main advantage is that stellar and
calibration light follow identical paths \citep[see
also][]{1982A&A...114..357B, 2010A&A...515A.106F}, which also allows using gas
absorption lines for establishing a precise wavelength scale and specification
of the spectrograph instrumental line shape over the entire spectral range
\citep[e.g., iodine;][]{1992PASP..104..270M, 1995PASP..107..966V,
  1996PASP..108..500B}. For reference, a laboratory spectrum is used, which is
obtained with an FTS at a much higher resolution and signal-to-noise ratio
(S/N) than the \'echelle spectra.

Over the last half century, the field of precision Doppler experiments
has developed into an industry, with many new spectrographs at a variety of
facilities. So far, frequency calibration is typically limited at the
m\,s$^{-1}$ level
\citep[$\varv/c = \Delta f/f = \Delta \lambda/\lambda \sim 10^{-8}$;
see][]{2016PASP..128f6001F} or slightly better in individual targets
\citep[e.g.,][]{2020A&A...639A..77S}. Calibration strategies generally fall
into two categories \citep{2010exop.book...27L}, known as the {iodine
  cell technique} (see above) and the {simultaneous reference technique}
\citep{1996A&AS..119..373B}. In this work, we present a strategy for using an FTS
to establish the accurate frequency scale for any calibration
spectrum. This can be used to create calibration spectra optimized in shape
and coverage for astronomical spectrographs, and accurately referenced
across the entire spectral range. To demonstrate this, we employ a model of
molecular iodine absorption, and show that the model can be used either
instead of an observed template for the iodine cell technique, or as
a simultaneous reference if illuminated with a flatfield lamp.

\section{Methods}

\subsection{Fourier Transform Spectrometer}

An FTS records an interferogram produced by a Michelson interferometer with
one movable mirror \citep[other technologies exist; see,
e.g.,][]{2007ApJ...661..616H}. They are standard tools in laboratory
spectroscopy. Frequency calibration is achieved through a calibration laser
that provides a reference for the position of the movable mirror. In contrast
to grating spectrographs, the frequency scale in an FTS is, to very high
degree, linear in wavenumber because it is defined by interference phenomena
inherent to the instrument \citep{1972ifts.book.....B, Learner1988}. The only
free parameter in the frequency scale is the offset between the control laser
and the science light, which is typically known with an uncertainty of around
100\,m\,s$^{-1}$ Doppler shift. This is in stark contrast to grating
spectrographs, where the frequency of every individual pixel comes with a
substantial uncertainty. In practice, however, the optical path difference
between control laser and science light in an FTS can depend on frequency,
and is caused for example by dispersion in the beam splitter. In the complex spectrum
reconstructed from the interferogram, this causes a phase shift that varies
with frequency and needs to be corrected for. Phase errors can cause
significant frequency offsets between different parts of the spectrum, which
is why empirical verification of frequency linearity is important. The
phase shift is expected to be a relatively slowly varying function of
frequency, which is why a small symmetric portion of the interferogram is
sufficient for phase correction \citep[e.g., using the ``Mertz''
method;][]{mertz1965transformations, MERTZ196717, Learner:95}. We refer to
\citet{2001ftsp.book.....D} and \citet{2007ftis.book.....G} for details about
phase correction.

Another advantage of Fourier Transform Spectrometry is that the observed
interferogram analytically defines the spectrum as a sum of continuous
trigonometric functions. In other words, the sampling of the interferogram
does not translate into a spectrum sampled at a finite number of pixels, but
the spectrum can (in principle) be computed at arbitrary positions from the
interferogram. This allows arbitrarily high sampling of the spectrum and a
clean definition of the instrumental line shape; while sparse sampling is a
limiting factor for the measurement of spectral lines in astronomical grating
spectra, this problem does not exist in Fourier Transform
Spectrometry. Therefore, the latter provides a testbed for high-accuracy line
profile measurements.

The FTS offset can be determined from a calibration standard in the observed
spectrum \citep[see, e.g.,][]{2016A&A...587A..65R}; one of the main advantages of this approach
is that calibration features do not need to cover the same frequencies as the
science spectrum (from the Sun or other sources) because the interferometer
simultaneously receives information about the entire spectral range during a
scan \citep{1972ifts.book.....B}. We can therefore use one part of the
spectral range for calibration and another for the science spectrum.

Our setup is a commercial Bruker 125HR with a HeNe laser for reference. The
maximum optical path difference is 208\,cm, of which 47\,cm are symmetric
around the interferogram zero-point. We are using custom software for
computation of the spectrum, in particular for phase correction, which is
critical for our high-resolution spectra. We apply no apodization and we use
the Mertz method for phase correction \citep[][see also
Appendix\,\ref{sect:phase}]{2001ftsp.book.....D}.

For this work, we used the VIS setup of our evacuated FTS covering the
spectral range 10,000--25,000\,cm$^{-1}$ (4000--10,000\,\AA). For the
iodine-LFC measurements, we combined the light of the two sources using a
dichroic beamsplitter with a 6800\,\AA\ cutoff wavelength outside the FTS and
coupled the combined light into the FTS input fiber \citep[see
also][]{2020spie11447e..3qs,2020spie11447e..a9s}. The input fiber is
hexagonal in shape, which ensures good near- and far-field scrambling of the two light
sources.

\subsection{Laser frequency comb}

\begin{figure*}
  \centering
  \mbox{
    \parbox{.95\textwidth}{
      \resizebox{.95\textwidth}{!}{\includegraphics{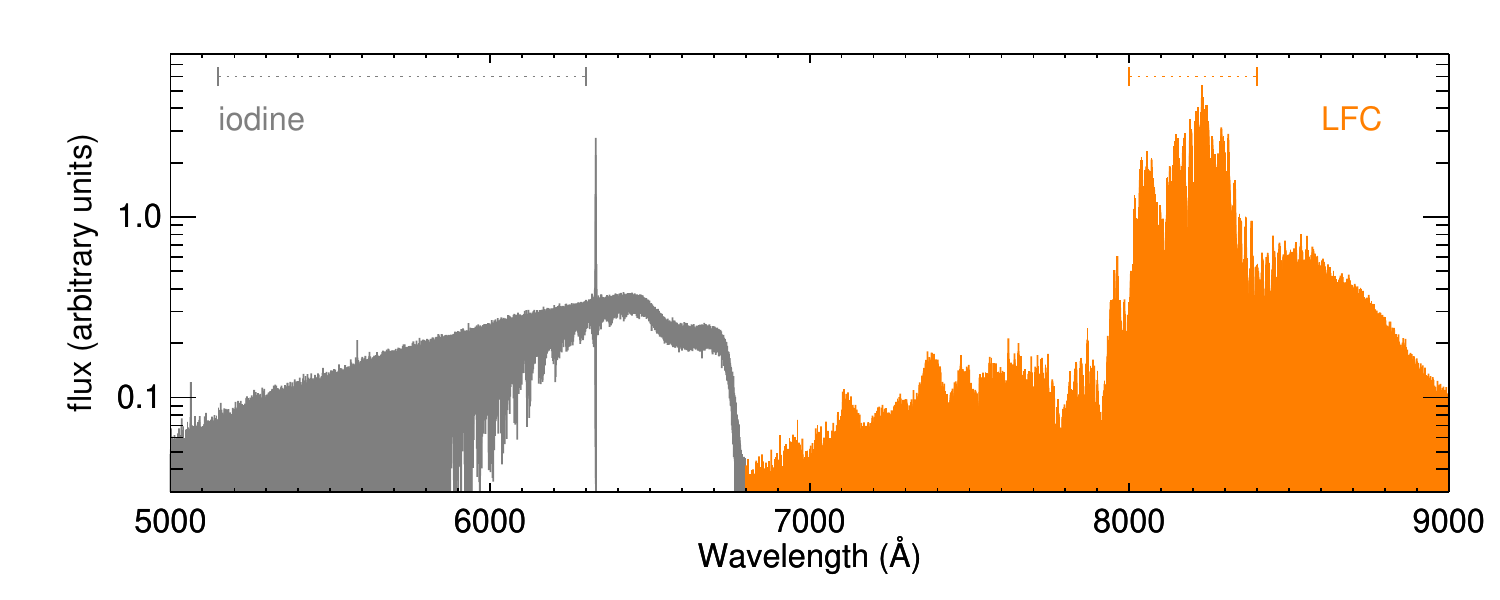}}
    }}
  \caption{\label{fig:I2_LFClines}Spectrum of I$_2$ absorption and the LFC as obtained
with the FTS. Light of the two sources is combined with a dichroic beam
splitter; at wavelengths shorter than 6800\,\AA, only light from the I$_2$
absorption enters the spectrometer (gray). At longer wavelengths, the spectrum
includes only the LFC spectrum (orange). The I$_2$ spectrum follows the
intensity of the illuminating lamp and shows the molecular absorption band
structure with only very weak lines at long wavelengths beyond 6600\,\AA\
($T \approx 44^{\circ}$C). The feature at 6330\,\AA\ is the HeNe reference
laser of the FTS, and the dip at 6500\,\AA\ is caused by the transmission curve of
the beam splitter. The LFC spectrum is highly dynamic and provides lines
across the entire wavelength range and particularly strong lines in the range
8000--8400\,\AA. Wavelength ranges used for Doppler measurements in the I$_2$
and LFC spectra are indicated with dashed lines at the top.}
\end{figure*}

The frequency spectrum of an LFC is a broadband comb of equidistant emission
lines, which can be stabilized with high accuracy and precision
\citep{Reichert1999,Diddams2000,Jones2000}. The position of each line is
determined by two degrees of freedom: the repetition rate, $f_\mathrm{rep}$, and the
carrier-envelope-offset frequency, $f_\mathrm{CEO}$, governed by the relation
\begin{equation}
\label{eq:lfc}
  f_{n, \mathrm{LFC}} = f_\mathrm{CEO} + n f_\mathrm{rep},
\end{equation}
in frequency units, or $\lambda_{n, \mathrm{LFC}} = c/f_{n, \mathrm{LFC}}$ in
units of wavelength. Accuracy and precision are achieved by phase locking
$f_\mathrm{rep}$ and $f_\mathrm{CEO}$ to a stable reference oscillator, such
as an atomic clock. LFCs are important tools in metrology and, among many
other applications \citep{Diddams2010,Picque2019}, are promising calibration
light sources for astronomy
\citep{Herr2019,Probst2020,2021A&A...646A.144S}. In the following, we will use
the expression "absolute calibrator" to indicate that the calibration is
traced back to the definition of second as much as the actual setup will
allow.

The LFC in our setup is a LaserQuantum (Novanta) taccor comb. The source laser
is a pulsed Ti:sapphire laser with a repetition rate of approximately
1\,GHz. The commercial setup includes an f-2f interferometer for measuring the
offset frequency.  Two frequency generators (Rohde \& Schwarz SMB100A and
HMF2500) are used for locking $f_\mathrm{rep}$ and $f_\mathrm{CEO}$, respectively. Our time
base reference is a GPS-8 from MenloSystems with a precision and accuracy
of better than $10^{-12}$ in one second, which provides a 10\,MHz signal for both
frequency generators. For the measurements in this work, the two degrees of
freedom of the LFC were stabilized to $f_\mathrm{rep} = 1.0019850000$\,GHz and
$f_\mathrm{CEO}= 377.4000000$\,MHz.

To generate a supercontinuum, we use a photonic crystal fiber stub of 14.5\,mm
in length (NL-2.8-850-02) tapered by Vytran according to specifications retrieved
from simulations based on the approach of \cite{ravi2018}. Our simulation
approach is designed to generate a stable supercontinuum, as detailed in
\cite{Debus2021}.

\subsection{Absorption cell with molecular I$_2$}

Our iodine absorption cell setup consists of off-the-shelf components from
Thorlabs: The iodine cell is a GC19100-I, and the heater assembly is a GCH25-75
controlled with a TC200-EC unit. We use a fiber-coupled tungsten-halogen lamp
(HL-2000-HP-FHSA, OceanOptics) with OAP fiber couplers (RC08FC-P01, Thorlabs)
to guide light through the cell. The whole optical assembly is wrapped in
aluminum foil and placed in a styrofoam insulated box for additional
temperature stability. Typically, the temperature is stable to within 100\,mK
on the timescale of one day.

\subsection{Model of molecular iodine absorption}
\label{sec:i2model}

The model we use for molecular iodine absorption is based on a description of
the rovibronic structure of the I$_2$ B-X spectrum calculated from molecular
potentials for the two electronic states and their hyperfine parameters
informed by high-precision measurements of the B-X spectrum of I$_2$ in the
visible \citep{knoeckel2004}. Depending on the temperature assumed in the
modeling, relative intensities of individual transitions are predicted. The
expected accuracy of the transition frequencies is better than 3\,MHz
($\sim$2\,m\,s$^{-1}$) in the wavelength range of 5260--6670\,\AA. From the line
list, we construct a model spectrum through broadening each individual
spectral line according to temperature Doppler broadening and the FTS
instrument profile.

For the measurement used in this work, we computed the line list according to
an I$_2$ temperature of $T = 44^{\circ}$C. The model spectrum contains a total
number of 4,427,241 lines in the range of 5150--6300\,\AA. We included all lines
in our model regardless of their predicted absorption intensity. The I$_2$
spectrum contains on average between 30 and 100 lines per 1\,km\,s$^{-1}$
Doppler width, which is 100--300 lines in one resolution element seen by a
typical astronomical spectrograph ($R = 100,000$). We scale all line
intensities from the model calculations by a factor of 3800 to approximately
match our observed spectrum.

\section{Observational data and frequency calibration}

\subsection{LFC and I$_2$ spectra}

We combined light from an LFC and an I$_2$ absorption cell and simultaneously
obtained their spectra in the same FTS scan. A dichroic beam splitter
separated the light such that the FTS received light at wavelengths of shorter
than 6800\,\AA\ only from the I$_2$ cell and at longer wavelengths only from
the LFC. We obtained 19 spectra with ten scans each on May 25, 2023; the total scan
time per spectrum was 22\,minutes. One example spectrum with both components
is shown in Fig.\,\ref{fig:I2_LFClines}.

The simultaneous observation of I$_2$ and the LFC spectrum allows a direct
comparison between the LFC line positions and the absolute I$_2$ line
frequencies. From the information about the LFC line position, we determined
the zero-point offset of the FTS frequency solution establishing an accurate
frequency scale for the I$_2$ spectrum.

\subsection{Absolute frequency calibration from LFC spectrum}
\label{sect:LFCzero}

\begin{figure}
  \centering
  \resizebox{.9\linewidth}{!}{\includegraphics[viewport=10 10 700 810, clip=]{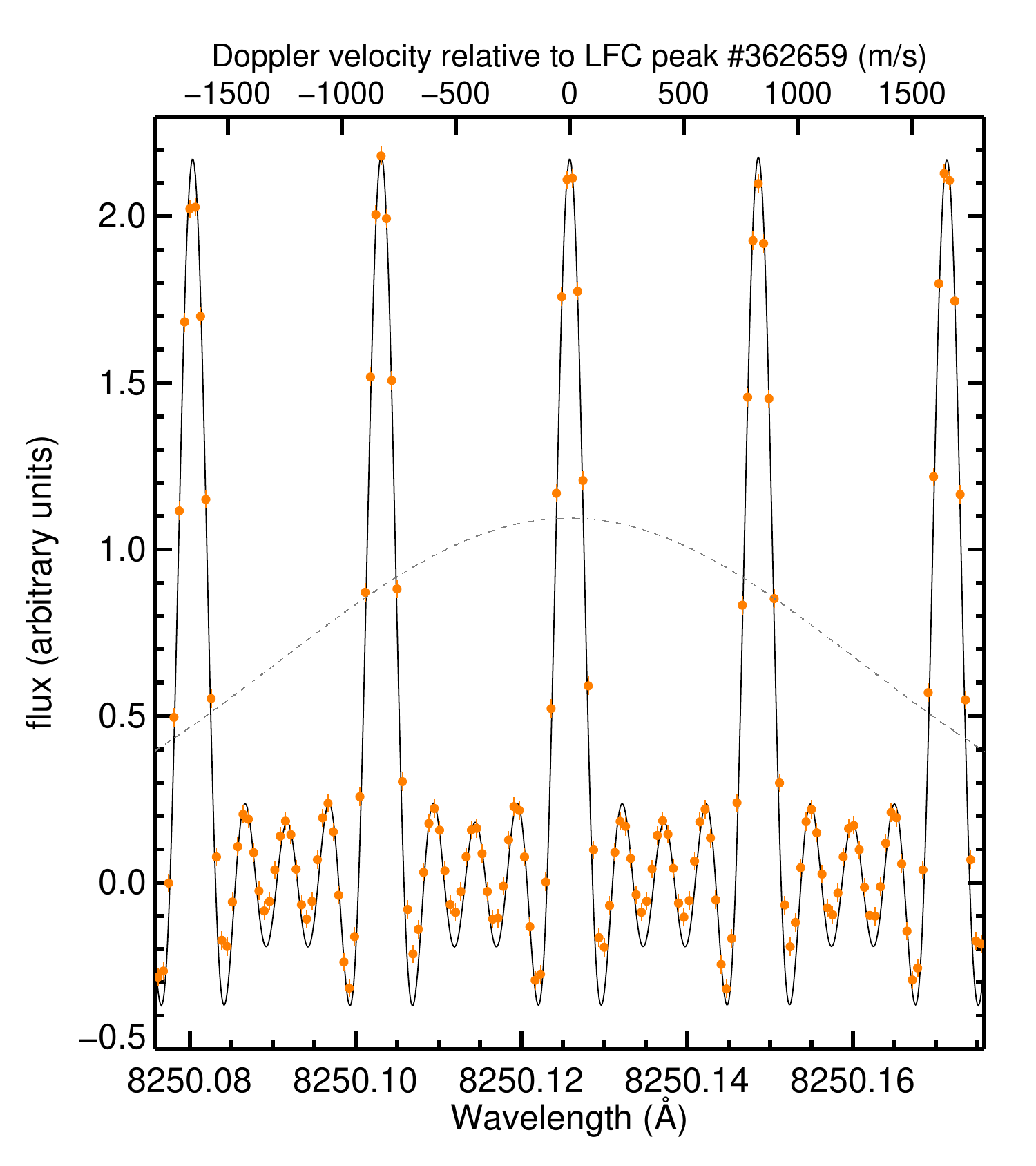}}
  \caption{\label{fig:LFClines}Zoom onto a spectrum of the LFC as observed with
    the FTS (orange circles with vertical bars for the estimated uncertainty)
    compared to the analytical description of the FTS instrumental line shape
    convolved with the LFC peak pattern around the comb tooth $n = 362,659$
    (solid line). For comparison, the instrumental line shape of a
    spectrograph with a resolution of $R = 100,000$ is shown (dashed line).}
\end{figure}

We aim to determine the zero-point of our combined I$_2$ and LFC spectra from
the region of the spectra containing the strongest LFC lines
(Fig.\,\ref{fig:I2_LFClines}). We show a small portion of one LFC spectrum in
Fig.\,\ref{fig:LFClines} that we use as example for the 19 spectra
incorporated in the final analysis. The width of the individual lines is
determined by the maximum optical path difference, $L$. We used $L = 136$\,cm
as a compromise between spectral resolution and scan time.  We compare the
spectral region around the LFC peak $n = 362,659$ to an analytical model of the
instrumental line shape convolved with a model LFC peak pattern. The
instrumental line shape is a sinc-function with a width determined by the
maximum optical path difference. Additional broadening caused by the
finite-sized aperture is introduced by a convolution with a box function of
width $1/2L$ \citep{2001ftsp.book.....D}. Assuming optimum aperture, we
estimated the effective resolution as the quadratic sum of the effects from
finite scan length, $L$, and corresponding aperture;
$R = k/\Delta k_L = \sqrt{2}Lk \approx 2.3 \cdot 10^6$ at wavenumber
$k = 12121$\,cm$^{-1}$ ($\lambda = 8250$\,\AA) and using
$\Delta k_L = 1/\sqrt{2}L$. We did not attempt any shape optimization to
account for optical imperfections.

The observed spectrum computed from the interferogram is shown with
uncertainties estimated from the spectral noise determined in the spectral
range of 6600--6700\,\AA,\ which is almost completely free of spectral features. The
root mean square (rms)\ noise is $\sigma_{\rm rms} = 0.012$ in the flux units shown in
Fig.\,\ref{fig:I2_LFClines}. The rms\ noise is independent of the pixel
sampling of our spectrum and represents the case of uncorrelated sampling only
if the number of pixels, $N$, is equal to the number of resolution elements,
$N_L$; that is, if the wavenumber stepsize is $\Delta k = \Delta k_L$, with
wavenumber $k = 1/\lambda$ and $\lambda$ the wavelength. To scale the noise
level to our oversampled spectrum, we computed the noise following the scaling
relation
\begin{equation}
  \sigma^2 =
  \sigma_{\rm rms}^2 \cdot N/N_L =
  \sigma_{\rm rms}^2 \cdot \Delta k_L / \Delta k =
  \sigma_{\rm rms}^2 / \sqrt{2} L\Delta k,
\end{equation}
where $\Delta k$ is the stepsize in the spectrum used.

For the LFC spectrum in Fig.\,\ref{fig:LFClines}, absolute peak positions are
known from Eq.\,(\ref{eq:lfc}), and peak intensity is adjusted manually to
visually match the observed spectrum; we did not attempt to perform a formal
fit here. The only additional free parameter of the observed spectrum is the
global zero-point offset.

The analytical description of the instrumentally broadened LFC spectrum
matches the (offset corrected) observed spectrum to a high degree. The
individual LFC peaks show a clear sinc pattern in which the sidelobes
partially overlap between the peaks, producing a periodic pattern that is
significantly different from noise. We refrain from a detailed analysis of the
FTS instrumental line shape but note a slight asymmetry between the depths of
the blue and red minima visible around the main peaks; for example, the red
minima are less deep and do not fully extend to the expected depth. For
comparison, we include in Fig.\,\ref{fig:LFClines} an instrumental line shape
according to a resolution of $R = 100,000$, the typical resolution of
astronomical high-resolution spectrographs (dashed line). Very often, in
astronomical spectra, such a resolution element is sampled with an element of
approximately~3 pixels per resolution. This underlines the potential of the
Fourier transform technique to offer an improved understanding of calibration
source characterization and line center fitting.

\begin{figure*}
  \centering
  \mbox{
    \parbox{.95\linewidth}{
      \resizebox{.56\linewidth}{!}{\includegraphics[viewport=30 05 630 450, clip=]{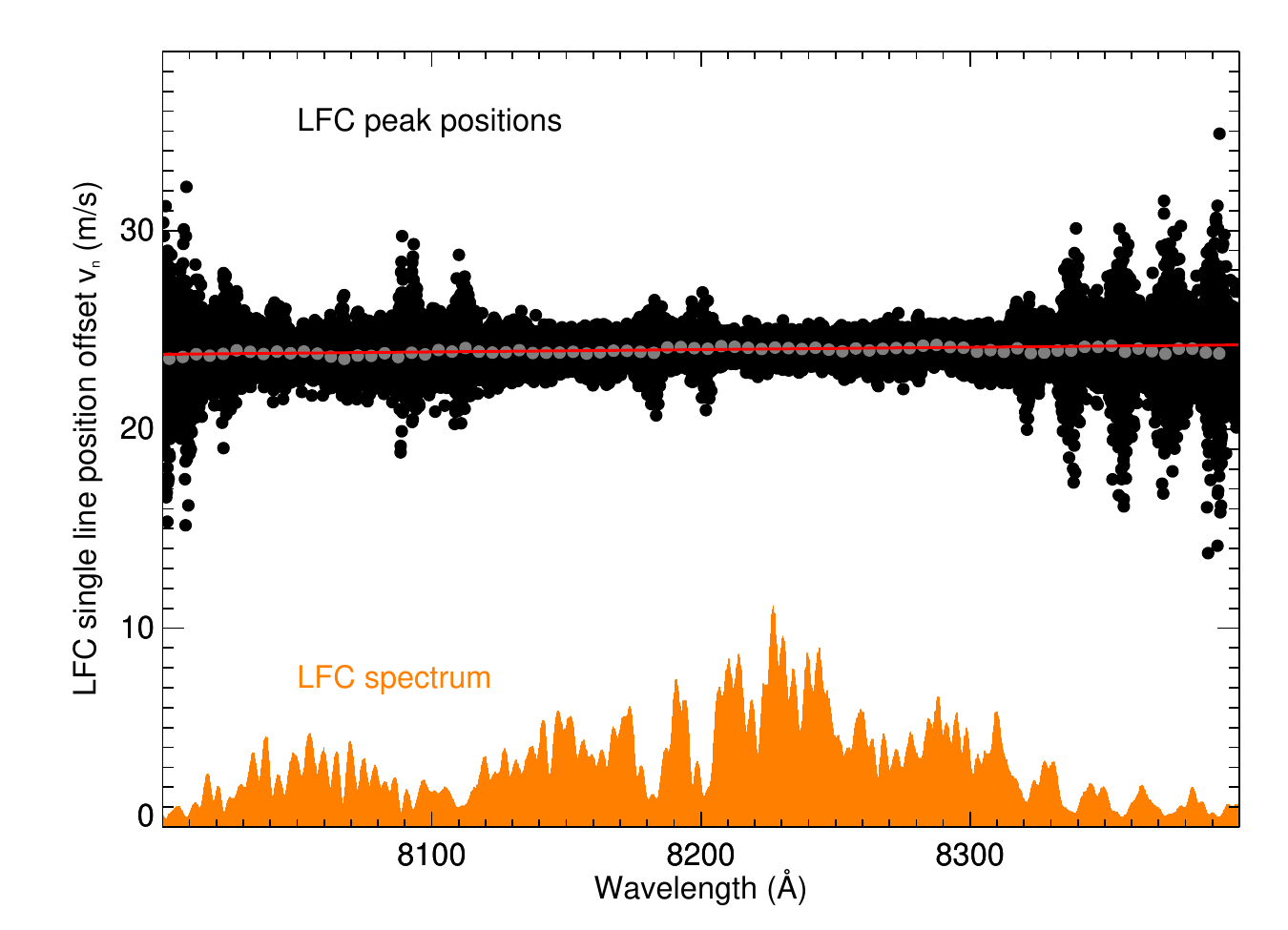}}
      \resizebox{.43\linewidth}{!}{\includegraphics{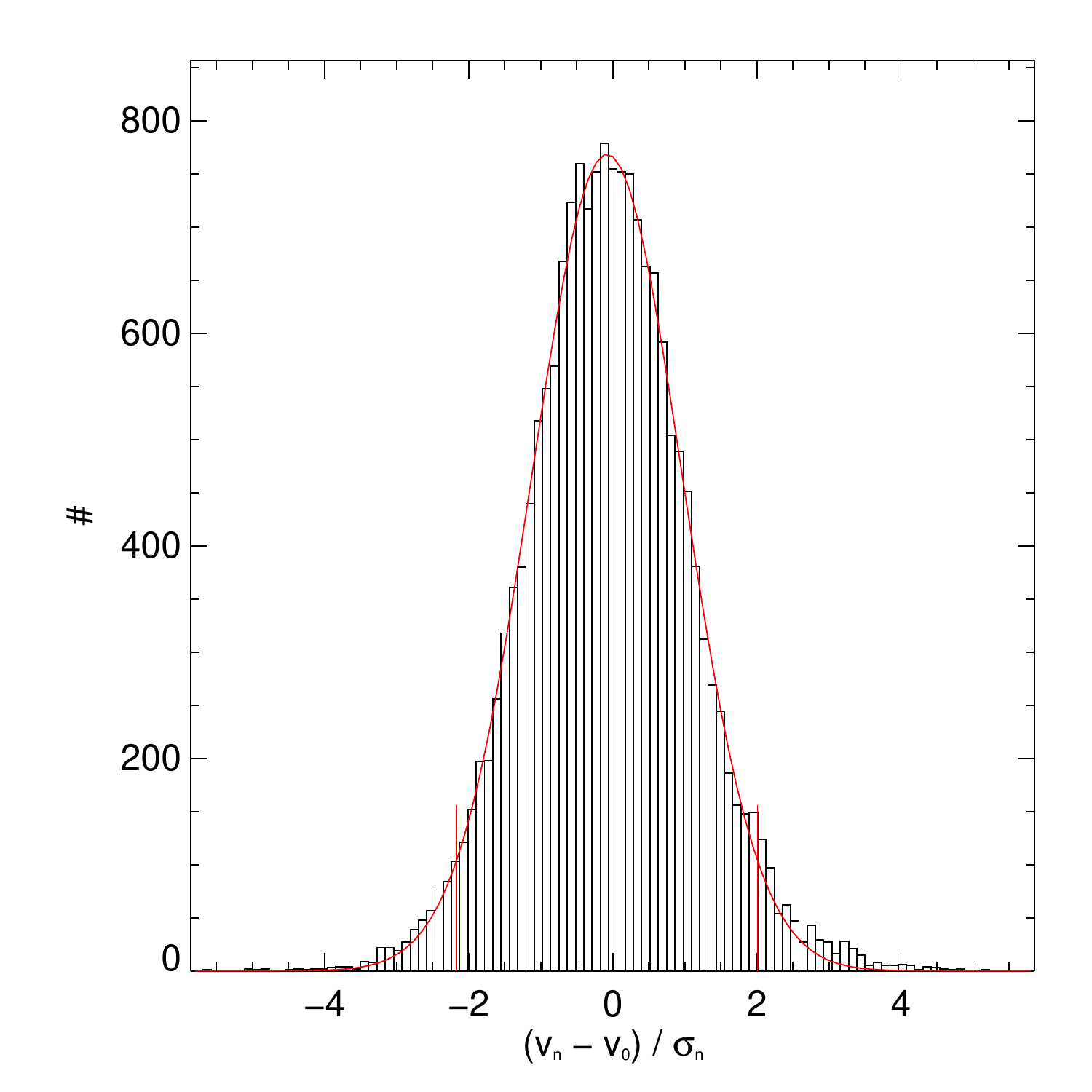}}
    }}
  \caption{\label{fig:LFCzeros}Measurements of LFC line
    positions. \emph{Left:} Measurements of 17,807 individual LFC peak
    positions, $\varv_n$ (black circles). Weighted means for 
bins
 of 5\,\AA\ in width    are shown as gray circles. A linear fit to all measurements is shown as
    a red line; the small slope determined is not visible on this scale. The
    spectrum of the LFC is overplotted (orange) in linear arbitrary units. The
    width of the LFC peak position distribution is wider in the areas of lower
    signal. \emph{Right:} Histogram of residuals between observed and true LFC
    positions, $\varv_n$, after subtracting the FTS zero point, $\varv_0$ and
    normalized by the uncertainty of the measured line position,
    $\sigma_n$. The red curve indicates a Gaussian distribution with a
    2$\sigma$ interval indicated by the red vertical bars. The distribution is
    only 5\,\% broader than expected from uncertainties, $\sigma_n$.}
\end{figure*}

We determined the zero point (i.e., the offset to the control laser in the
FTS) of our observation, $\varv_{0}$, from 17,807 individual LFC peaks in the
wavelength range of 8000--8400\,\AA. For every known LFC peak,
$\lambda_{n, \mathrm{LFC}}$ in Eq.\,(\ref{eq:lfc}), we fit a sinc function to
the observed spectrum in a window of $250$\,m\,s$^{-1}$ in Doppler width around the
peak center, which approximately covers the main peak of the instrumental line
shape. Fit parameters were the amplitude of the LFC peak, the velocity offset
between the peak center observed in the spectrum ($\lambda_{n, \mathrm{FTS}}$)
and its true position ($\lambda_{n, \mathrm{LFC}}$),
$\varv_n = c~(\lambda_{n, \mathrm{FTS}} - \lambda_{n,
  \mathrm{LFC}})/\lambda_{n, \mathrm{LFC}}$, and the width of the sinc
function. We ignored the additional (symmetric) broadening from the finite
aperture applied in the FTS because its main effect is an increase in the peak
width. The individual line positions were fit with an uncertainty depending on
peak flux; 69\,\% and
34\,\% of the lines showed uncertainties of below 1\,m\,s$^{-1}$ and below 50\,cm\,s$^{-1}$, respectively.

The distribution of individual LFC peak position measurements across the
spectral range is shown in the left panel of Fig.\,\ref{fig:LFCzeros}. Peak
position measurements are scattered around a mean value with a distribution
that is wider at regions of lower peak amplitude, which is consistent with the
assumption that lines with higher intensity are better determined. We fit a
linear slope to the velocity offsets, $\varv_{n}$, to determine the zero point
of our I$_2$ and LFC spectrum, $\varv_{0}$, and to search for a potential
slope in the wavelength solution, $s$, using the linear model
\begin{equation}
  \label{eq:vn}
  \varv_n = \varv_{0} + s \cdot (\lambda_n - \lambda_\mathrm{c}),
\end{equation}
with $\lambda_\mathrm{c} = 8200$\,\AA.

In the spectrum shown in Fig.\,\ref{fig:I2_LFClines}, we find a zero-point
offset of $\varv_0 = 23.945 \pm 0.004$\,m\,s$^{-1}$, that is, the statistical
uncertainty of the mean wavelength accuracy of our spectrum is
$\sigma_{\varv_0} = 4$\,mm\,s$^{-1}$, which is representative of all 19
spectra. The slope we determine from the fit is
$s = -0.49 \pm 0.05$\,mm\,s$^{-1}$\,\AA$^{-1}$, which is a (formally)
statistically significant slope of around\ 20\,cm\,s$^{-1}$ over the range of
400\,\AA. We experimented with different wavelength ranges, finding that the
value of the slope scatters around zero depending on the choice of range.
From this we conclude that the slope value is dominated by systematic rather
than statistical uncertainties and that our results are consistent with zero
slope or smaller than $|s| = 1$\,mm\,s$^{-1}$\,\AA$^{-1}$ (approximately\
$3\cdot10^{-8}$ lin.\ dispersion). This is in agreement with the results in
\cite{huke2019characterization}, where we found the linear dispersion to be
below $10^{-8}$ at wavelengths of 8000--9800\,\AA. Following the same argument,
we estimate that the uncertainty in the zero-point determination is
approximately 1\,cm\,s$^{-1}$, which is slightly larger than the formal fit
result because of systematic effects.

To assess whether the fit positions were influenced by systematic effects, we
investigated the distribution of $\varv_n$ around $\varv_0$, divided by their
fit uncertainty $\sigma_n$. This distribution is shown in the right panel of
Fig.\,\ref{fig:LFCzeros}. It is consistent with a Gaussian distribution with a
width that is only 5\,\% larger than expected from the uncertainties, which is
an indication of a realistic noise estimate. To search for
wavelength-dependent patterns in the offset, we overplot the weighted means in bins of
5\,\AA\ in width in the left panel of Fig.\,\ref{fig:LFCzeros} (gray
circles). These values scatter around the overall mean with a standard
deviation of 16\,cm\,s$^{-1}$ without evidence for clear systematic uncertainties. From
this, we can rule out systematic patterns of several tens of\,angstroms\ in length and
exceeding a few 10\,cm\,s$^{-1}$ in the wavelength range of 8000--8400\,\AA.  We
note that hidden systematic uncertainties can be caused by the fact that we ignore the
additional finite-aperture broadening, imperfections of the instrumental line
shape, blends caused by the far wings of the instrumental line shape, or by
spectral features like water-absorption, which are not taken into account.

\section{Analysis of model I$_2$ absorption spectrum} 
\label{sect:I2analysis}

\subsection{Absolute frequencies}

With the zero-point determination, we established an accurate frequency
solution for our combined I$_2$ and LFC spectra, which means that the
frequency solutions of our iodine spectra were calibrated with an overall
uncertainty on the cm\,s$^{-1}$ level. We compared our observed spectra to
synthetic spectra based on the model of \cite{knoeckel2004} and determined
remaining Doppler velocity offsets between model and observations,
$\Delta$RV. To see whether the offset showed any dependence on frequency, we
performed a fit of the I$_2$ model spectrum to our observations in 302
spectral chunks of 200\,km\,s$^{-1}$ width across the wavelength range of
5150--6300\,\AA. The chunk size is a compromise between velocity uncertainty
per chunk and the spectral resolution of our analysis. We computed the fits for
each spectrum after zero-point correction with the LFC and averaged the
results from 19 spectra. For each chunk, fit parameters were the Doppler
velocity offset between the model and the observed spectrum, a scaling
parameter for the I$_2$ line absorption intensity, and a linear slope for
continuum normalization. Uncertainties of the offsets for each chunk per
single spectrum, as calculated from the spectrum noise, were below
2\,m\,s$^{-1}$ for 32\,\% and below 4\,m\,s$^{-1}$ for 90\,\% of the
$19 \cdot 302 = 5738$ individual computations. After averaging over the 19
exposures, the median uncertainty per chunk is 0.54\,m\,s$^{-1}$ with 93\,\%
of the values below 1\,m\,s$^{-1}$. This allows us to investigate the accuracy
of the model's absolute frequency scale and its dependence on wavelength.

\begin{figure}
  \centering
  \resizebox{\linewidth}{!}{\includegraphics[viewport=10 30 690 690, clip=]{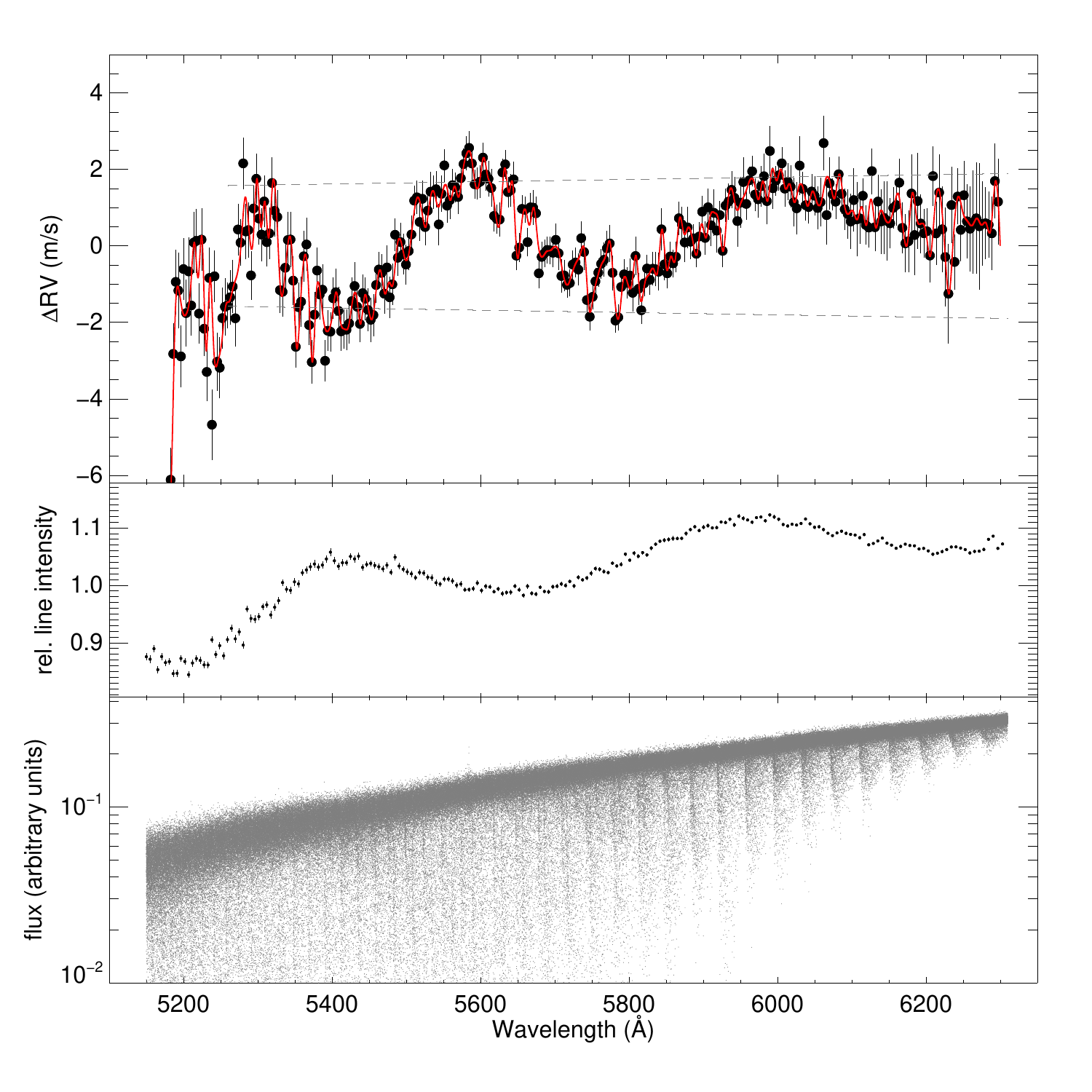}}
  \caption{\label{fig:I2shift}Doppler velocity offset between the I$_2$ model
    and observations after zero-point correction. \emph{Top:} Individual
    Doppler offsets in chunks of 200\,km\,s$^{-1}$ width (black circles). Gray
    dashed lines show uncertainty limits as estimated in
    \cite{knoeckel2004}. The red line shows a spline fit to the Doppler
    offsets showing a systematic pattern in the offsets. \emph{Middle:}
    Relative I$_2$ line intensities as determined from the fitting procedure
    (see text). \emph{Bottom:} Observed spectrum of I$_2$ for comparison.}
\end{figure}

The absolute Doppler velocity offsets, $\Delta$RV, between the I$_2$ model
spectrum and our observations are shown in Fig.\,\ref{fig:I2shift}. The model
frequencies are centered around zero Doppler offset: the frequencies of the
LFC-corrected I$_2$ spectra accurately coincide with the model
predictions. Specifically, Doppler offsets are distributed within the
estimated frequency uncertainty pattern \citep{knoeckel2004}, which is
indicated as gray dashed lines in the top panel of Fig.\,\ref{fig:I2shift}.
We therefore conclude that the I$_2$ model accurately predicts the I$_2$
absorption line frequencies, and that the model frequencies are useful for an
{absolute} calibration of astronomical spectrographs.

We allow a scaling of the absorption line intensity because we suspect that
the calculated transition intensities applying the Franck-Condon principle
show deviations from observations that depend on frequency. The middle panel
of Fig.\,\ref{fig:I2shift} shows that the intensities of the vibronic
transitions indeed deviate by more than 10\,\% from the average. We confirmed
that the systematic variation of the transition frequencies shown in the top
panel of Fig.\,\ref{fig:I2shift} are independent of line intensity by
carrying out the same analysis but keeping line intensity constant over all
frequencies. This clearly indicates that improvements in the modeling of the
spectra should mainly focus on the energy scale (potential functions and
hyperfine interaction) and not the intensity (relaxing the Franck-Condon
principle).

The pattern in $\Delta$RV consists of one long-period wave overlaid by a
rapidly oscillating pattern that coincindes with the I$_2$ absorption band
structure. Uncertainties in $\Delta$RV for each chunk are shown in
Fig.\,\ref{fig:I2shift} and demonstrate that the long-period wave and also
parts of the oscillating pattern are significantly different from random
noise. The amplitude of the long-period wave is approximately 2\,m\,s$^{-1}$,
which significantly exceeds the frequency linearity determined from the
LFC lines in Fig.\,\ref{fig:LFCzeros}. While the LFC lines cover a
different frequency range, we see no obvious reason why the linearity should
be very different in the frequency range used here. A potential source of
systematic errors in determinations of radial velocity is the phase correction. We show the
reconstructed phase for one of our spectra in Appendix\,\ref{sect:phase}, in
which we find no evidence for a systematic pattern resembling the $\Delta$RV
signature. To verify the robustness of the $\Delta$RV pattern, we computed
$\Delta$RV using the power spectrum (instead of the phase-corrected spectrum)
and found that the $\Delta$RV pattern also appears. This demonstrates that
phase errors are an unlikely source of the $\Delta$RV pattern.

We therefore believe that the pattern of Doppler offsets is dominated by
systematic shifts in the model frequencies that vary through the rotational
bands of the I$_2$ B-X spectrum. We suggest that our observations be used to
improve the Doppler offset in the model spectrum for this systematic
effect. To visualize and interpolate the offset pattern, we show a spline fit
to the velocity offsets after applying a smoothing (red line in top panel of
Fig.\,\ref{fig:I2shift}). The spline represents the effective frequency
correction required to adjust the model spectrum. This correction provides a
frequency solution that is {accurate} to within 0.5--1\,m\,s$^{-1}$ across
the range of 5300--6150\,\AA. We emphasize that this value does not correspond to
the average performance for the full wavelength range but applies to every
chunk fitted in the spectrum.

\subsection{Comparison to the high-$S/N$ observation}

\begin{figure*}
  \centering
  \mbox{
    \parbox{.9\linewidth}{
      \resizebox{.9\linewidth}{!}{\includegraphics[viewport=30 10 700 385, clip=]{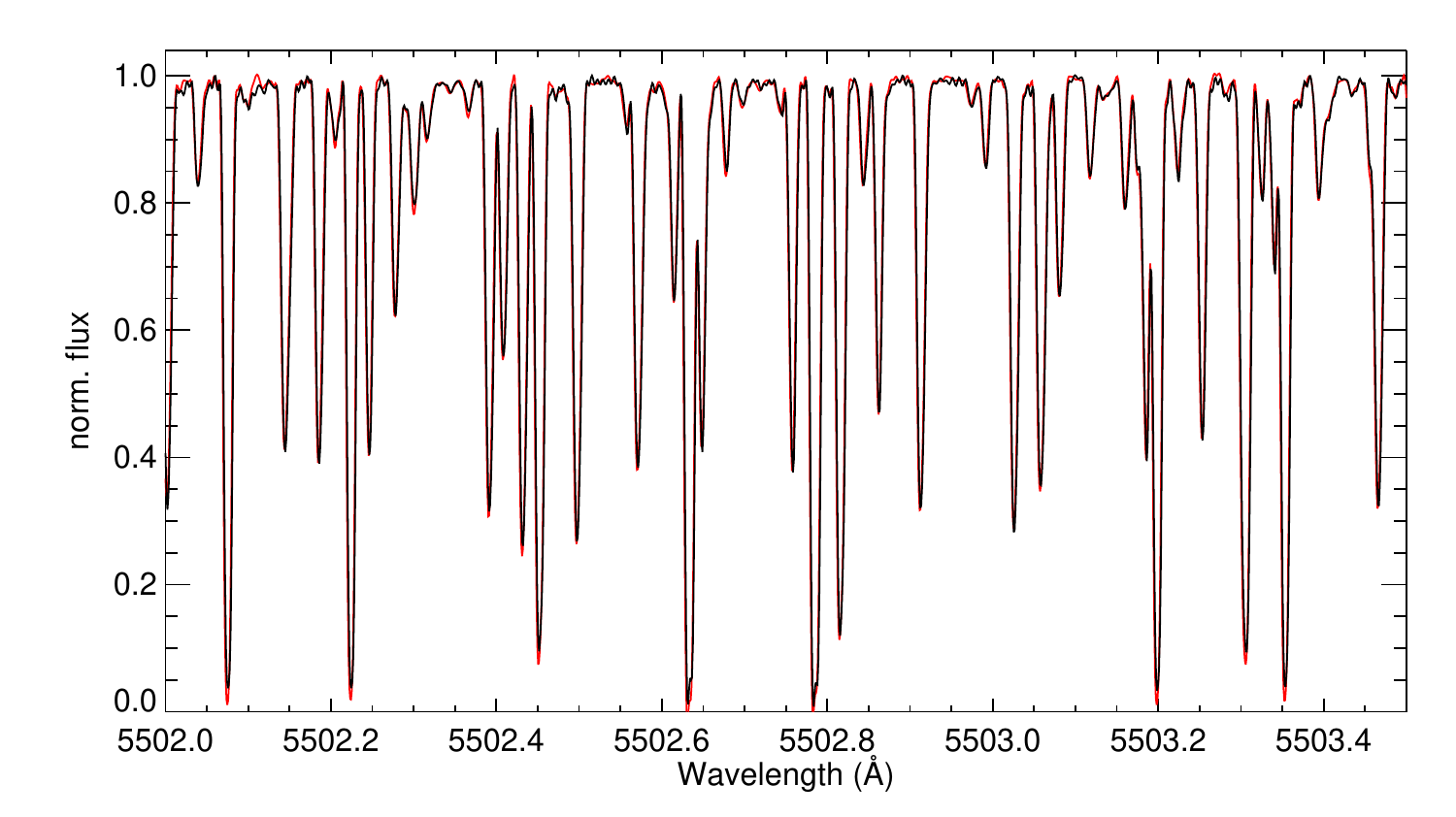}}
    }}
  \caption{\label{fig:I2_model}Comparison between a deep (92\,h) FTS iodine
    absorption spectrum (black, scan length $L = 45$\,cm) and a model spectrum
    (red) for $T=44^{\circ}$C.}
\end{figure*}

To complement our comparison between the I$_2$ model and observations, we
obtained a deep spectrum of I$_2$ absorption with our FTS. We added 227
individual observations with ten scans each taken Jan 20--24, 2023. Here, we
used the symmetric configuration of our FTS with a maximum optical path
difference of $L = 45$\,cm, applying forward-backward scanning. Each scan took
approximately 24\,min, with a total scan time of 92\,h. Co-addition was performed
without individual zero-point correction because typical shifts of a few
m\,s$^{-1}$ between consecutive observations are not relevant for our visual
comparison. We show the spectrum in the wavelength range of 5502--5504\,\AA\ in
Fig.\,\ref{fig:I2_model} together with a model for $T = 44^{\circ}$C. The S/N
of the co-added spectrum in this wavelength range is approximately 2400 at a
resolution $R = 1.2 \cdot 10^6$. The model provides an excellent fit even at
this resolution, reproducing essentially all I$_2$ lines. We particularly
emphasize the quality at regions where strong lines overlap, which is a good
diagnostic of line depths and spectral resolution. We also note that the model
reveals some areas with very few lines where an almost clean continuum is
visible. In these regions, the data show a ripple pattern that we believe is
not caused by noise but by the sinc function instrumental line shape. Our
model includes the instrumental line shape but we did not attempt to
empirically determine the line shape, nor did we search for any missing lines
or lines underestimated in the model spectrum.

\section{Discussion}

\subsection{Absolute cross calibration with an FTS}

The linearity of the FTS frequency scale allowed us to project the
frequency accuracy of the  LFC onto the wavelength range of the I$_2$ spectrum. The
simultaneous (dichroic) observation of LFC and I$_2$ spectra at different
wavelengths provides I$_2$ spectra on a frequency scale that is linear and
accurate with an offset uncertainty of smaller than
$\Delta \varv_0 = 1$\,cm\,s$^{-1}$ ($\Delta \varv_0 / c = 3 \cdot
10^{-11}$). In general, an FTS can project the accuracy of an absolute
frequency standard, for example, an LFC or an I$_2$ spectrum, into wavelength ranges
not originally covered by the frequency standard. The projection of frequency
accuracy can be carried out on the spectrum from any light source, in
particular wavelength reference spectra optimized for astronomical
spectrographs, such as an FP.

Such reference spectra can then be used to calibrate astronomical
spectrographs in close analogy to the strategy followed when employing an LFC
but without the need for the LFC light to cover the entire wavelength range or
enter the spectrograph. This significantly relaxes requirements on bandwidth,
free spectral range, and peak uniformity (albeit peak stability should not
vary strongly on timescales of the FTS scan). For example, the comb teeth of a
1\,GHz LFC as used in our setup can be adequately distinguished at 8500\,\AA\
by an FTS with a maximum pathf olarger than 30\,cm, and the
resolution of such an FTS at 6000\,\AA, $R = 10^{6}$, is sufficient for the
characterization of the reference spectrum.

Furthermore, all calibration sources can be characterized at spectral
resolutions far exceeding that of the astronomical spectrograph, and
can be monitored for variability. This can lift the paradoxical situation whereby
spectra from calibration sources are never seen with any other instrument than
the one being calibrated.

\subsection{Calibration concept for astronomical spectrographs}

% \textbf{Frequency calibration of astronomical spectrographs is realized through
%   external calibration light sources for which the spectral pattern must be
%   known. The LFC is regarded an ideal calibrator because it provides dense
%   sampling of lines, narrow features allowing reconstruction of the
%   line-spread function (LSF), and, most importantly, because each individual
%   spectral feature is known to great accuracy (better than $3\cdot10^{-11}$ or
%   1\,cm\,s$^{-1}$). Alternative calibrators providing absolute frequency
%   information are hollow-cathode lamps, for example ThAr or UNe, but their
%   individual lines are not known better than $3\cdot10^{-8}$ or 10\,m\,s$^{-1}$,
%   and their position can vary during operation and with lifetime. Secondary
%   calibrators providing limited absolute but very precise relative information
%   are FPs. They can be tailored for the spectral format (frequency range,
%   resolution) of the spectrograph, but individual line positions cannot be
%   predicted and vary in time.} 

In practice, high-resolution spectrographs are calibrated using a suite of
calibration sources. Spectra of an LFC or hollow-cathode lamps are taken once
or a few times per day providing information about the absolute positions of
wavelengths on the detector. In addition, a stable FP is often used to
interpolate between hollow-cathode lamp lines \citep{2010SPIE.7735E..4XW,
  2015A&A...581A.117B}, or extrapolate to wavelengths not covered by the other
sources. Some observatories also use the FP simultaneously during science
observations to track the spectrograph drift.

\begin{figure}
  \centering
  \mbox{
    \parbox{.99\linewidth}{
%      \resizebox{.99\linewidth}{!}{\includegraphics[viewport=5 60 710 490, clip=]{DichroicReference.pdf}}
      \resizebox{.99\linewidth}{!}{\includegraphics[viewport=30 50 695 450, clip=]{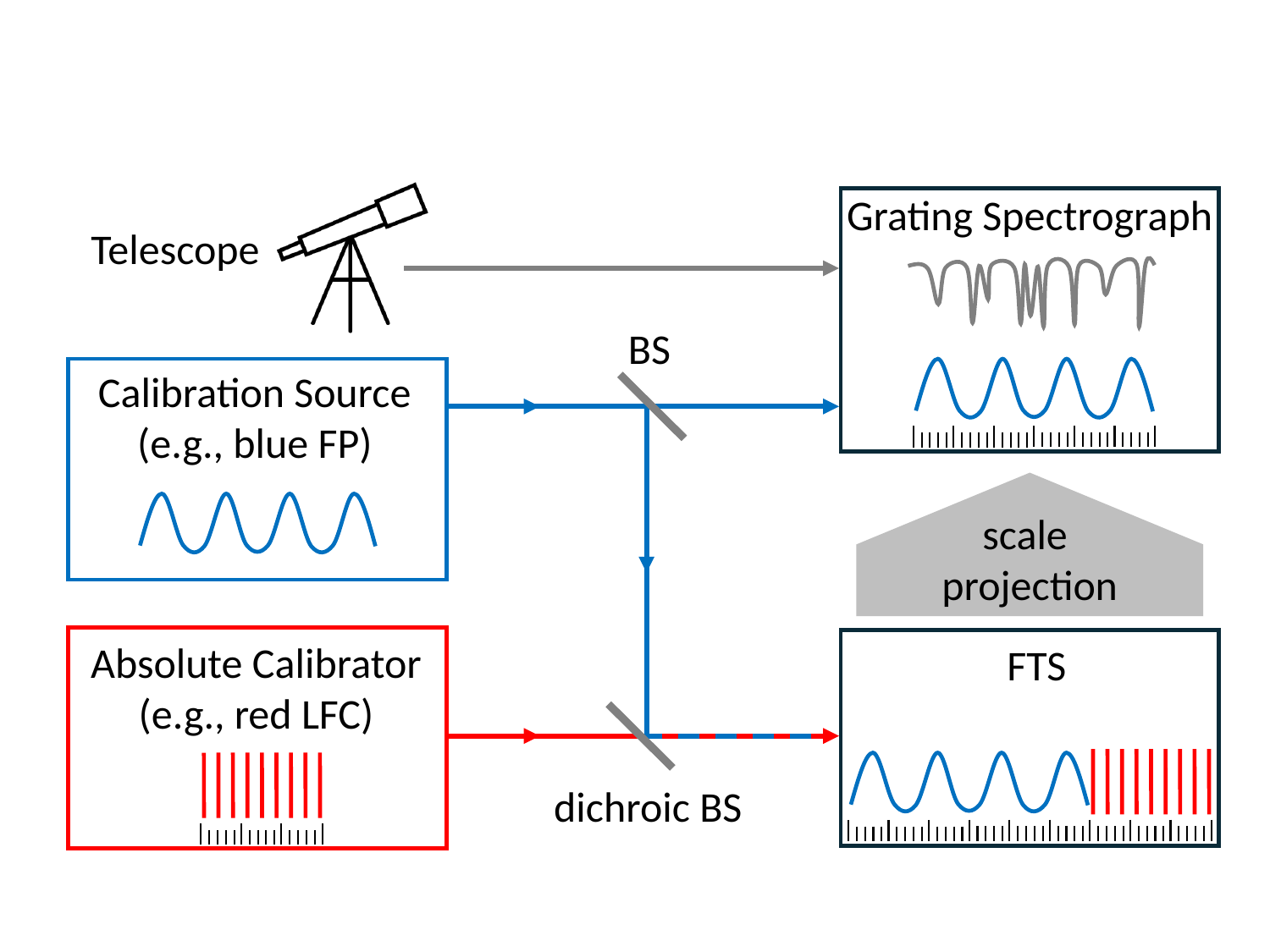}}
    }}
  \caption{\label{fig:CalConcept}Dichroic calibration concept. The FTS is used
    to project the frequency scale from the absolute calibrator (shown in red)
    onto the spectrum of the calibration source (blue), which is used for
    calibrating the grating spectrograph.}
\end{figure}

We suggest that an FTS can be used to employ any light source in order to provide
{accurate} information about the wavelength solution
(Fig.\,\ref{fig:CalConcept}). In practice, one would use an absolute standard
---such as an LFC--- to calibrate the FTS in a limited spectral range outside
the range covered by the astronomical spectrograph (e.g., 8000--8400\,\AA).
  %Because of the higher
  %spectral resolution of the FTS, the LFC can have relatively narrow mode
  %spacing, e.g., 1\,GHz, and is required to cover only a relatively small
  %frequency range.
During each calibration exposure of the astronomical spectrograph, a fraction
of the light from the calibration source needs to be channeled into the FTS
where a high-resolution spectrum with an accurate frequency solution is
obtained in the full spectral range relevant for the astronomical
observations, such as 3800--8000\,\AA. Thus, for every individual exposure taken
with the astronomical spectrograph, the FTS provides an accurate frequency
solution for any of the calibration sources.

The availability of accurate spectra for each calibration exposure massively
relaxes requirements on calibration source stability. For example, with this
strategy, an FP only needs to be stable for the duration of the observation
because variability becomes visible at the significantly higher spectral
resolution of the FTS. We argue that the full characterization of reference
spectra during the time of each observation removes critical free parameters
during the calibration process that are otherwise not accessible ---such as variability in line strengths from hollow cathode lamps or an LFC--- and
that the opportunity to use {any} type of calibration source can lead to
superior calibration strategies and reliability. For example, a tunable FP
could be used to iteratively cover all detector pixels in the astronomical
spectrograph if accurate calibration information is available.

For calibration of the FTS offset, an LFC is a viable choice. Alternatively, a
stabilized laser could be sufficient, and we demonstrate in the present paper that
I$_2$ also provides offset accuracy at the sub-m\,s$^{-1}$ level. The actual
choice of calibration source and strategy depends on the individual setup
but is very flexible. In our solar observatory in G\"ottingen, we are
obtaining spectra of the Sun in the range of 4000--6800\,\AA\ with our FTS. The
FTS could be calibrated with the LFC but we avoid using the latter for
everyday observations for practical reasons. Instead, we calibrate the
instrument with simultaneous FP measurements in the wavelength range of
6800--9000\,\AA. The FP itself is calibrated every day using a simultaneous
observation of the FP with I$_2$ at 4000--6800\,\AA\
\citep{10.1117/1.JATIS.9.4.045003}.

The design of a full calibration plan exceeds the scope of this paper and
depends on the actual setup and requirements of the astronomical
spectrograph. It is probably realistic to cover the full wavelength range of
astronomical spectrographs with one FP and use the FTS spectra to provide
sub-m\,s$^{-1}$ accuracy. A tunable FP can further improve the wavelength
solution while high-finesse FPs could be used to provide narrow emission
lines useful for characterizing the instrumental profile over the entire
frequency range. The performance of this critical step with respect to an LFC
remains to be tested. With the relaxed requirements on absolute calibration
and the flexible choice of dichroic beamsplitters, technical solutions are
available for a wide range of applications, including visual and infrared
spectrographs.

\subsection{Iodine as an absolute calibrator}

We demonstrate in Section\,\ref{sect:I2analysis} that spectra computed from
the molecular potential model from \citet{knoeckel2004} describe observations
taken with an I$_2$ absorption cell to a high degree, and that the modeled and
corrected frequencies are accurate to a level of below 1\,m\,s$^{-1}$.  Thus,
iodine cells are useful absolute calibrators, and it is possible to obtain
their spectra and wavelength information from (1) FTS measurements or (2) I$_2$
model spectra.

For calibrating astronomical spectrographs, an iodine cell, illuminated by a
flatfield lamp, can provide an economic solution in the wavelength range of
5200--6300\,\AA. This can be useful in addition to other calibration sources, such as hollow cathode lamps and Fabry-P\'erot etalons, which are affordable for
small-budget observatories. At observatories that include LFCs in their
calibration plan, an iodine cell can provide an additional calibration
source. Its spectrum more closely resembles stellar absorption spectra and is
therefore useful for investigating the impact of potential differences in the
trace profiles between emission (LFC/Fabry-P\'erot) and absorption spectra
\citep{2021A&A...646A.144S}. Most importantly, an iodine cell can be operated in
the telescope beam, while observing bright stars  for example. Thus, accurate iodine
cell spectra are useful for verifying Doppler offsets potentially caused from
using different light paths for calibration and science light.

Applications involving I$_2$ absorption cell spectra usually rely on reference
spectra obtained with an FTS at a significantly higher resolution than used in
the astronomical data. Accurate model spectra could be used instead, with the advantage of this being that model spectra are absolutely noise free and
can be computed at arbitrary spectral resolution and cell temperature. For
example, the model can allow the cell temperature and line
intensity (or partial pressure) to be fitted, which  can potentially help to reduce
requirements on temperature stability and I$_2$ condensation issues. We would
always strongly recommend to obtain high-quality reference FTS spectra for any
iodine cell used for astronomical spectroscopy. Nevertheless, the flexibility
offered by a model spectrum, in addition to the FTS scan, can hardly be
overrated as long as the model sufficiently matches the real spectrum.

Our results could also have a very important impact on molecular spectroscopy.
To the best of our knowledge, such a complete simulation of a molecular
spectrum in connection with an observation has never been achieved over a range of
5200 to 6200\,\AA\ with convincing consistency. Our approach can easily be adapted to
the modeling of observed molecular spectra.

\section{Summary}

We obtained simultaneous observations of an LFC and an I$_2$ absorption cell
at different wavelengths in an FTS. LFC lines in the wavelength range of
8000--8400\,\AA\ were used to test the consistency and linearity of the FTS
frequency solution, and to determine offset and dispersion. The dispersion is
found to be below 1\,m\,s$^{-1}$ per 1000\,\AA, and the offset is determined with
an accuracy of 1\,cm\,s$^{-1}$. Comparison between the observed spectrum and an
analytic model of the instrumental line shape demonstrates exquisite spectral
quality at a resolution some 20 times higher than astronomical high-resolution
spectra. This allows the development and study of spectral analysis algorithms at
the sub-m\,s$^{-1}$ level without systematic limitations related to spectral
quality.

All individual spectra were zero-point corrected with the information from the
LFC lines, and the I$_2$ wavelength range of 5150--6300\,\AA\ was used to compare
observations against model I$_2$ spectra. The comparison was carried out in
spectral chunks of 200\,km\,s$^{-1}$ width to search for systematic
variability in the model line frequencies. The offsets show a characteristic
pattern that we attribute to the I$_2$ band structure. The pattern is
centered around zero velocity and is consistent with the absolute frequency
uncertainty estimated in the model ($\la 2$\,m\,s$^{-1}$). This means that the
model spectra are useful for providing an absolute frequency scale in I$_2$
broadband spectra, for example when illuminating an iodine cell with a flatfield
lamp or starlight. We argue that the systematic frequency pattern can be
corrected from our comparison between model and FTS observations, and that the
final frequency scale is {accurate} within an uncertainty of
1\,m\,s$^{-1}$ across the wavelength range of 5200--6400\,\AA. The high
consistency between model and observations opens the opportunity to use the
high flexibility of model spectra for analysis involving I$_2$ absorption
lines.

The high accuracy demonstrated in I$_2$ and LFC spectrometry shows the
potential of using an FTS for calibrating astronomical spectrographs. The
possibility of obtaining a referenced spectrum from any light source with an
FTS allows the spectra of any calibration source to be obtained with an absolute
frequency scale known with an uncertainty of better than 10\,cm\,s$^{-1}$. We
argue that simultaneous FTS monitoring of calibration sources alleviates the need
for an absolute calibrator to cover the entire frequency range. The FTS can
extend the frequency accuracy from a small wavelength portion over its entire
range. This also allows great flexibility in the design of calibration
sources, which will help improve the performance of calibration strategies
for the next generation of high-precision Doppler experiments in astronomy.

\begin{acknowledgements}
  We regret to notify that Horst Knöckel passed away very recently July
  2024. He was the main contributor over many years on the spectroscopic work
  and development of the iodine model, on which our present work is based. He
  would be delighted to see these fruits of his work. We acknowledge the help
  of J.\,Dabrunz with the extraction of line lists from the
  \texttt{IodineSpec} program, and we thank the anonymous referee for a
  \emph{very} helpful report. We thank T.\,Schmidt, F.\,Kerber, L.\,Pasquini,
  P.\,Huke, and members of the ELT Working Group \emph{Line Calibration} for
  discussions about results and applications. M.\,Debus was funded through the
  Bundesministerium f\"ur Bildung und Forschung (ELT-ANDES, 05A2023).
\end{acknowledgements}

\bibliographystyle{aa}
\bibliography{refs}

\begin{appendix}

  \section{Phase correction}
  \label{sect:phase}

  In Fourier Transform Spectrometry, the spectrum is computed through Fourier
  transform of the measured interferogram. In theory, the interferogram is
  symmetric around zero mirror displacement, and Fourier transform of the real
  and symmetric interferogram leads to a real spectrum with no imaginary part,
  $S(k)$. In practice, however, the interferogram contains noise, its zero
  point can not be exactly determined and is a function of wavelength, which
  all leads to a spectrum with complex components,
  \begin{equation}
    C(k) = R(k) + i I(k).
  \end{equation}
  This \emph{phase error} can be corrected under the assumption that the
  spectrum is real (the imaginary part is zero). The dependence between $C(k)$
  and $S(k)$ can be represented by multiplication with a phase that depends on
  wavenumber, $k$,
  \begin{equation}
    C(k) = S(k) \exp{(i \Phi(k))}.
  \end{equation}
  With $C(k)$ computed from the measured interferogram, one way to reconstruct
  $S(k)$ is multiplicative phase correction, an algorithm that follows the
  strategy developed by \citet{mertz1965transformations, MERTZ196717}. With
  \begin{equation}
    \Phi(k) = \arctan{\left(\frac{I(k)}{R(k)}\right)},
  \end{equation}
  we can write
  \begin{eqnarray}
    S(k) & = & C(k) \exp{(-i \Phi(i))}\\
         & = & R(k) \cos{(\Phi(k))} + I(k) \sin{(\Phi(k))},
  \end{eqnarray}
  which is the phase-corrected, real spectrum. We refer to \citet{Learner:95}
  and \citet{2001ftsp.book.....D} for more detailed descriptions of phase
  correction. Wrong phase correction can lead to a significant deformation of
  spectral features and, in particular, to apparent Doppler shifts
  proportional to the phase error.
  
\begin{figure}[h]
    \centering
    \resizebox{\linewidth}{!}{\includegraphics{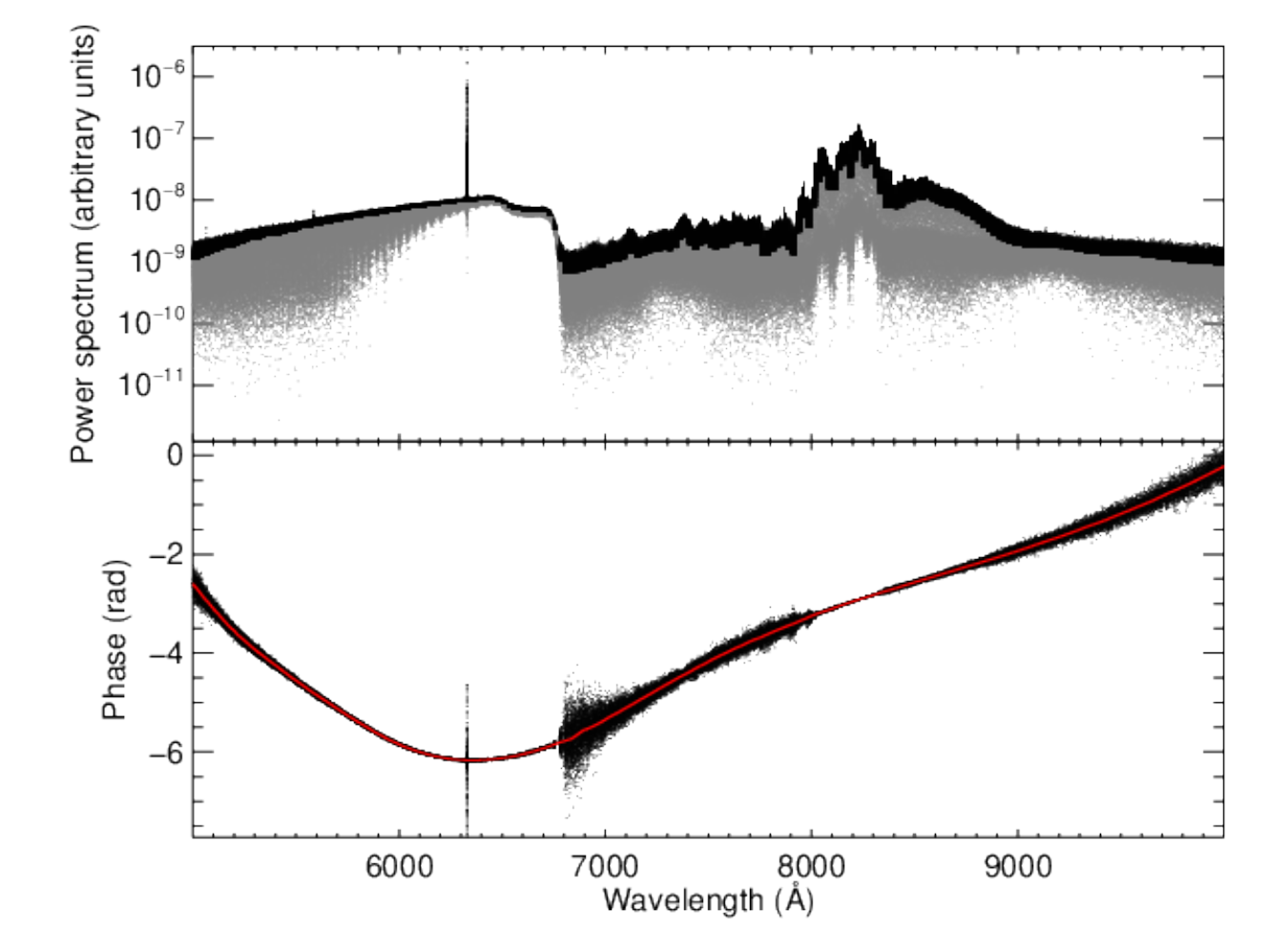}} % use this: convert Phase.eps Phase.pdf
    \caption{\label{fig:phase}\emph{Top:}Power spectrum of the combined iodine
      and LFC light as in Fig.\,\ref{fig:I2_LFClines}. The top 20\,\% of the
      data points are used for phase correction and shown in
      black. \emph{Bottom:} Phase reconstructed from the symmetric part of the
      interferogram following the 'Mertz' algorithm. The red line shows a
      spline fit to a running median of the phase data. This smooth curve is
      used for phase correction.}
  \end{figure}

  For one of our spectra containing I$_2$ and the LFC signal (as in
  Fig.\,\ref{fig:I2_LFClines}), we show the phase in
  Fig.\,\ref{fig:phase}. The phase is derived from the data points with
  highest intensity. For small portions of the spectrum, we select the 20\,\%
  of the points that show the highest intensity (black points in top panel of
  Fig.\,\ref{fig:phase}), and fit a spline curve through their phase, which is
  shown as red curve in the bottom panel of Fig.\,\ref{fig:phase}.

  For phase correction, we used the 47\,cm long symmetric part of the FTS. We
  estimated that at a systematic displacement of 1\,m\,s$^{-1}$ in spectral
  features could be caused by a phase error of approximately 0.01\,rad
  \citep[see][]{2001ftsp.book.....D}. A pattern as the one shown in
  Fig.\,\ref{fig:I2shift} could be caused by a similar pattern in phase with
  such an amplitude. The data we used for phase determination is distributed
  symmetrically around the smoothed curve with a 1-$\sigma$ width of
  approximately 0.01\,rad, and the statistical uncertainty of the phase per
  10\,\AA\ bin is $\sim 0.001$\,rad or 10\,cm\,s$^{-1}$. From this we can
  conclude that a phase pattern in frequency of 1\,m\,s$^{-1}$ amplitude would
  be detectable in our data. As a consistency check, we tested our RV
  determination using the power spectrum instead of the phase-corrected
  spectrum. We found the same $\Delta$RV-pattern in this analysis, which
  confirms that the pattern is stable against errors in the phase
  correction. In conclusion, we believe that our analysis is robust against
  systematic errors on the scale of about 10\,cm\,s$^{-1}$.
  %That said, we emphasize that
  %we do not claim that our setup is free of additional systematics that we
  %either neglected or are unaware of, and that further characterization of
  %systematics will be useful.
  % For example, we tested the distribution of phase points around the mean
  % including points with lower intensity. The distribution shown signs of
  % asymmetry at wavelengths below 5800\,\AA. We do not know the physical
  % reason for this asymmetry.
   
\end{appendix}

\end{document}